\documentstyle[pre,aps,floats,epsf,amssymb]{revtex}
\begin{document}
\draft

\twocolumn[\hsize\textwidth\columnwidth\hsize\csname@twocolumnfalse%
\endcsname

\title{Self-organization of traffic jams in cities:
effects of stochastic dynamics and signal periods}


\author{Debashish Chowdhury$^{\dag}$ and Andreas Schadschneider} 

\address{Institute for Theoretical Physics, 
University of Cologne,
Z\"ulpicher Strasse 77,
D-50937 K\"oln, Germany} 

\maketitle 


\begin{abstract}
We propose a new cellular automata model for  
vehicular traffic in cities by combining (and appropriately modifying) 
ideas borrowed from the Biham-Middleton-Levine (BML) model of city 
traffic and the Nagel-Schreckenberg (NS) model of highway traffic. We 
demonstrate a phase transition from the "free-flowing" dynamical phase 
to the completely "jammed" phase at a vehicle density 
which depends on the time periods of the synchronized signals and the 
separation between them. The intrinsic stochasticity of the dynamics, 
which triggers the onset of jamming, is similar to that in the NS model, 
while the phenomenon of complete jamming through self-organization as 
well as the final jammed configurations are similar to those in the BML 
model. Using our new model, we have made the first investigation of the 
time-dependence of the average speeds of the cars in the 
"free-flowing" phase as well as the dependence of flux and jamming on 
the time period of the signals. 

\end{abstract}


Over the last half century various concepts and techniques of 
fluid dynamics and statistical mechanics have been successfully 
applied to understand several fundamental aspects of vehicular 
traffic flow ~\cite{p1,p3}. The "particle-hopping" models 
~\cite{p5,p6,p7} of vehicular traffic are usually formulated 
using the language of cellular automata (CA)\cite{p8}. These 
models are closely related to some of the microscopic models of 
driven systems of interacting particles, which are of current 
interest in non-equilibrium statistical mechanics\cite{p9}. 

A one-dimensional CA model of highway traffic and a two-dimensional 
CA model of city traffic were developed 
independently by Nagel and Schreckenberg (NS)\cite{p5} and Biham, 
Middleton and Levine (BML)\cite{p10}, respectively. Highway traffic 
becomes gradually more and more congested in the NS model with the 
increase of density. Traffic jams appear in the NS model because of 
the {\it intrinsic stochasticity} of the dynamics\cite{p6} but no  
jam persists for ever.
On the other hand, a first order phase 
transition takes place in the BML model at a finite non-vanishing 
density, where the average velocity of the vehicles vanishes 
discontinuously signalling complete jamming. In the BML model, the 
randomness, which is crucial for the jamming, arises only from the 
{\it random initial conditions}, as the dynamical rule for the 
movement of the vehicles is fully deterministic\cite{p10}. 

If each unit of discrete time interval in the BML model is 
interpreted as the time for which the traffic lights remain green 
(or red) before switching red (or green) simultaneously in a 
synchronized manner, then, over that time scale each vehicle, 
which faces a green signal, gets an opportunity to move from one 
crossing to the next. The generalization of the BML model that 
we propose here is, to our knowledge, the first attempt to describe 
explicitly the forward movement of the vehicles over smaller 
distances during shorter time intervals. We achieve this 
generalization by following the prescriptions of the NS model 
not only for describing the positions, speeds, accelerations 
and decelerations of the vehicles \cite{simon} but also for taking into 
account the interactions among the vehicles moving along the same lane 
of a street. We also modify some of the prescriptions of the BML model 
appropriately to take into account the signal-vehicle interactions 
and the interactions between vehicles approaching a crossing along 
different streets. 

Our main aim is to demonstrate that a phase transition from the 
"free-flowing" dynamical phase to the completely "jammed" phase 
takes place in our generalized model; the intrinsic stochasticity 
of the dynamics, which triggers the onset of jamming, is similar 
to that in the NS model, while the phenomenon of complete jamming 
through self-organization as well as the final jammed configurations 
are similar to those in the BML model. 

In the BML model a square lattice represents the network of the streets.
All the streets parallel to the $\hat{X}$-direction of a Cartesian
coordinate system are assumed to allow only single-lane
east-bound traffic while all those parallel to the $\hat{Y}$-direction
allow only single-lane north-bound traffic. Each of the lattice sites 
represents the crossing of a east-west street and a north-south street. 
In the initial state of the system, $N_x$ ($N_y$) vehicles are distributed 
among the east-bound (north-bound) streets.
The states of east-bound vehicles are updated in parallel at every odd 
discrete time step whereas those of the north-bound vehicles are updated 
in parallel at every even discrete time step following a rule which is 
a simple extension of the fully asymmetric simple exclusion 
process\cite{p9}: a vehicle moves forward by one lattice spacing if and 
only if the site in front is empty, otherwise the vehicle does not 
move at that time step. Jamming arises from 
the mutual blocking of the flows of east-bound and north-bound traffic 
at various different crossings. 
The BML model has been modified and 
extended\cite{p12,p13,p14,p15,p16,p17,p18,p19}.

We model the network of the streets as a $N \times N$ square lattice. 
The streets parallel to $X$ and $Y$ axes 
allow only east-bound and north-bound traffic, respectively, as in 
the original formulation of the BML model. A signal is installed at 
every site of this $N \times N$ square lattice where each of the 
sites represents a crossing of two mutually perpendicular streets. 
The separation between any two successive crossings on every street 
is assumed to consist of $D$ cells so that the total number of cells 
on every street is $L = N \times D$. The linear size of each cell 
may be interpreted as the typical length of a car; each of these 
cells can be either empty or occupied by at most one single vehicle 
at a time. Because of these cells, the network of the streets can be 
viewed as a decorated lattice. However, unlike the BML model\cite {p10} 
and the model of Horiguchi and Sakakibara\cite{p16}  
which correspond to $D = 1$ and $D = 2$, respectively, $D ~(< L)$ 
in our model is to be treated as a parameter. Note that $D$ introduces 
a new length scale into the problem.

The signals are synchronized in such a way that all the signals 
remain green for the east-bound vehicles (and simultaneously, red 
for the north-bound vehicles) for a time interval $T$ and then, 
simultaneously, all the signals turn red for the east-bound vehicles 
(and, green for the north-bound vehicles). Thus, the parameter $T$ 
introduces a new time scale into the problem.  

As in the original BML model, no turning of the vehicles is allowed. 
Therefore, the total number of vehicles on each street is determined 
by the initial condition, and does not change with time because of 
the periodic boundary conditions. 

Following the prescription of the NS model, we allow the speed $V$ 
of each vehicle to take one of the $V_{max}+1$ {\it integer} values 
$V=0,1,...,V_{max}$.
Suppose, $V_n$ is the speed of the $n$-th vehicle at time $t$ 
while moving either towards east or towards north. 
At each {\it discrete time} step $t \rightarrow t+1$, the arrangement of 
$N$ vehicles is updated {\it in parallel} according to the following 
"rules":

\noindent {\it Step 1: Acceleration.} If, $ V_n < V_{max}$, the 
speed of the $n$-th vehicle is increased by one, i.e., 
$V_n \rightarrow V_n+1$.

\noindent{\it Step 2: Deceleration (due to other vehicles or signal).} 
Suppose, $d_n$ is the gap in between the $n$-th vehicle and the 
vehicle in front of it, and $s_n$ is the distance between the same 
$n$-th vehicle and the closest signal in front of it. \\
{\sl Case I}: The signal is {\bf red} for the car under 
consideration:\\
If $min(d_n, s_n) \le V_n$ then $V_n \rightarrow min(d_n,s_n)-1$.\\
{\sl Case II}: The signal is {\bf green} for the vehicle under 
consideration:\\
There are two possibilities in this case: 
(i) When $d_n < s_n$, then $V_n \rightarrow d_n-1$ if $d_n \le V_n$, 
The motivation for this choice comes from the fact that, when 
$d_n < s_n$, the hindrance effect comes from the leading vehicle. \\ 
(ii) When, $d_n \geq s_n$, then $V_n \rightarrow min(V_n,d_n - 1)$   
if $min(V_n,d_n-1) \times \tau > s_n$, 
where $\tau$ is the number of the remaining time steps before the 
signal turns red. The motivation for this choice comes from the fact 
that, when $d_n \geq s_n$, the speed of the $n$-th vehicle at the next 
time step depends on whether or not the vehicle can cross the crossing 
in front before the signal for it turns red.  

\noindent{\it Step 3: Randomization.} If $V_n > 0$, the speed of the $n$-th 
vehicle is decreased randomly by unity (i.e., $V_n \rightarrow V_n-1$) with 
probability $p$ ($0 \leq p \leq 1$); $p$, the random deceleration probability, 
is identical for all the vehicles and does not change during the updating.

\noindent{\it Step 4: Vehicle movement.} Each vehicle is moved forward so 
that for the east-bound vehicles, $X_n \rightarrow  X_n + V_n$ 
where $X_n$ denotes the position of the $n$-th vehicle at time $t$  
while for the north-bound vehicles, $Y_n \rightarrow  Y_n + V_n$ 
where $Y_n$ denotes the position of the $n$-th vehicle at time $t$.

These rules are not merely a combination of the rules proposed by BML 
\cite{p10} and those introduced by NS \cite{p5} but also involves some 
modifications.
For example, unlike all the earlier BML-type models, a vehicle approaching 
a crossing can keep moving, even when the signal is red, until it reaches 
a site immediately in front of which there is either a halting vehicle or 
a crossing. Moreover, if $p=0$ every east-bound (north-bound) vehicle can 
adjust speed in the deceleration stage so as not to block the north-bound 
(east-bound) traffic when the signal is red for the east-bound 
(north-bound) vehicles. 

In our computer simulations, we begin with an initial configuration 
where $N_x$ and $N_y$ vehicles are put at random positions on the 
east-bound and north-bound streets, respectively. The states of the 
vehicles are updated in parallel following the rules mentioned above. 
After the initial transients die down, at every time step, we compute 
the average speeds $\langle V_x\rangle$ and $\langle V_y\rangle$ 
which are merely the averages of the instantaneous speeds of the 
east-bound and north-bound vehicles, respectively. The density 
$c = (N_x + N_y)/(2LN - N^2)$ of the vehicles is the ratio of the 
total number of cars and the total number of cells
in the system. Here we present the data for only the 
symmetric case $N_x = N_y$, for only a few set of values of the 
parameters $D, T, c, p, L, V_{max}$; more details will be published 
elsewhere~\cite{chowsh}. 

In the "free-flowing" phase of the BML model, both $\langle V_x\rangle$ 
and $\langle V_y\rangle$ 
oscillate between zero and a non-zero value 
periodically at odd and even time steps. In sharp contrast, the 
time-dependences of $\langle V_x\rangle$ and $\langle V_y\rangle$ are 
much more realistic in 
our model, as is evident from fig.\ \ref{fig1}.
Moreover, as expected, $V_g$, the maximum allowed values of 
$\langle V_x\rangle$ and $\langle V_y\rangle$ in the corresponding green 
phase, is smaller when the density $c$
is higher. In this parameter regime, following the 
switching of the red (green) signal to green (red), $\langle V_x\rangle$ rises 
(falls) to reach $V_g$ ($0$); the corresponding relaxation time is 
denoted by $t_g$ ($t_r$). For a given $c$, we now derive approximate 
analytical expressions for $t_g$ and $t_r$ in terms of $V_{NS}(c)$, 
the steady speed of the vehicles, for the vehicle density $c$, in 
the NS model with periodic boundary conditions. Then, using the 
numerical estimates of $V_{NS}(c)$ from computer simulations of the 
NS model we compute $\langle V_x\rangle$ and $\langle V_y\rangle$ for our
model and compare with the numerical data obtained from direct 
computer simulation.  

\begin{figure}[hbt]
\epsfxsize=\columnwidth\epsfbox{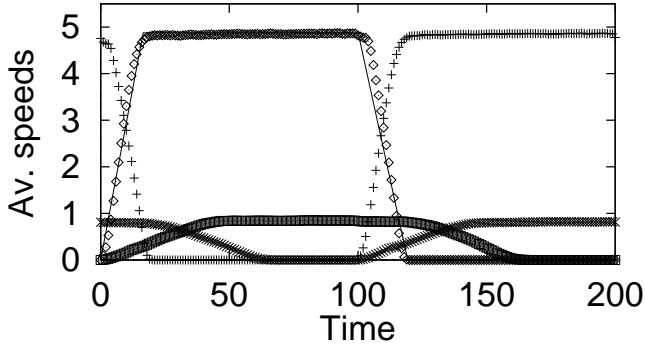}
\caption{Time-dependence of average speeds of vehicles. The symbols 
$\diamond$, $+$, $\square$ and $\times$ correspond, respectively, 
to $\langle V_x\rangle$ for $c = 0.1$, $\langle V_y\rangle$ for 
$c = 0.1$, $\langle V_x\rangle$ for $c = 0.5$ and $\langle V_y\rangle$ 
for $c = 0.5$. The common parameters are $V_{max} = 5, p = 0.1, 
D = 100$ and $T = 100$. The continuous line has been obtained from the 
equations (\ref{eq_Vg}) and (\ref{eq_Vr}).}
\label{fig1}
\end{figure}

We assume that during the red phase compact (i.e.\ without `holes') 
queues of length $N_q=cD$ are formed in front of each signal. We now 
estimate the time $t_g$ until the stationary speed $V_g = V_{NS}(c)$ 
is reached. There are two different additive contributions to $t_g$. 
First, the last vehicle in a compact queue of $N_q$ vehicles starts 
moving after $t_1=N_q/(1-p)$ timesteps since the leading vehicle in 
the remainder of the queue moves with probability $1-p$. Second, a 
halting vehicle reaches the speed $V_{NS}(c)$ after a time 
$t_2= V_{NS}(c)/(1-p)$ since it accelerates in each timestep with 
probability $1-p$. Thus,
\begin{equation}
t_g = t_1 + t_2 = \frac{cD+V_{NS}(c)}{1-p}
\label{def_tg}
\end{equation}
where we have assumed that the green phase starts at $t=0$.
Moreover, under the assumption that the relaxation is 
perfectly linear we obtain the following average speeds during a green 
phase:
\begin{equation}
V_g(t) = 
\cases{V_{NS}(c)t/t_g   & for $t < t_g$,\cr
  V_{NS}(c)             & for $t \geq t_g$,\cr}
\label{eq_Vg}
\end{equation}

In the stationary state of the green phase, there are, on the average, 
$\frac{1}{c}-1$ empty cells in front of each vehicle. So if the leading  
vehicle of a pair happens to be the last member of a queue already 
formed in front, then the following vehicle of that pair will move for  
$t_0=(\frac{1}{c}-1)/V_{NS}(c)$ with velocity $V_{NS}(c)$ and then stop 
suddenly since it reaches the tail of a queue. Therefore, on the 
average, it takes a time 
\begin{equation}
t_r = c D t_0 = \frac{(1-c)D}{V_{NS}(c)}.
\label{def_tr}
\end{equation}
to form a queue of length $N_q=cD$ after the first vehicle has stopped 
at the red signal. In general, the vehicle nearest to a signal will 
not stop immediately after the signal turns red, but will keep moving 
for some time, say, $t_{p}$. The distance of this vehicle from the 
signal is expected to be a fraction $\alpha$ of the average distance 
$\frac{1}{c}-1$ to the next vehicle ahead of it. Taking $\alpha = 1/2$, 
for example, one obtains $t_{p} = \frac{1-c}{2 c V_{NS}(c)}$. Hence, 
the average speed during a red phase starting at $t=0$ is given by 
\begin{equation}
V_r(t) = 
\cases{
  V_{NS}(c)                             & for $0 < t < t_{p}$,\cr
  V_{NS}(c)\left[1-\frac{t-t_{p}}{t_r}\right]
                                        & for $t_{p} < t < t_{p}+t_r$,\cr
  0                                     & for $t \geq t_{p}+t_r$.\cr}
\label{eq_Vr}
\end{equation}

For the validity of these estimates $T (\gg 1)$ should be sufficiently 
large to guarantee complete queueing of the vehicles during the red phase, 
$p$ should be small enough to ensure compactness of the queues, 
$c$ should be sufficiently small so that the vehicles emerging from a 
queue should not be hindered by the halting vehicles of another queue 
in front. Moreover, the smaller is $V_{max}$ the stronger is the 
deviation from the linear relaxation assumed above. Furthermore, we 
have assumed that all queues have the same length.

\begin{figure}[hbt]
\epsfxsize=\columnwidth\epsfbox{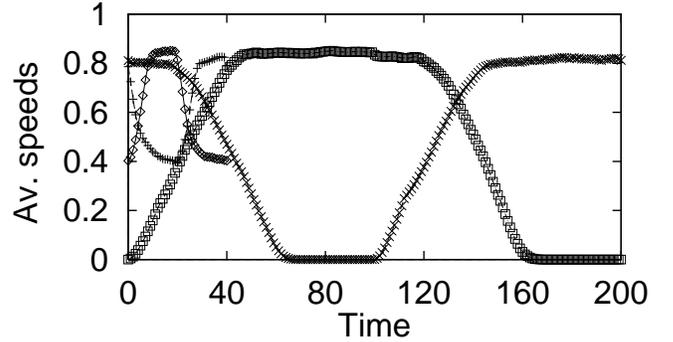}
\caption{Time dependence of average speeds of vehicles. The symbols 
$\diamond, +, \square$ and $\times$ correspond, respectively, to 
$\langle V_x\rangle$ for $T = 20$, $\langle V_y\rangle$ for $T = 20$, 
$\langle V_x\rangle$ for $T = 100$ and $\langle V_y\rangle$ for $T = 100$. 
The common parameters are $V_{max} = 5, p = 0.1, 
D = 100$ and $c = 0.5$. The data for $T = 20$ are shown only upto 
$40$ time steps to avoid overcrowding of data points. The 
lines are to serve as guides to the eye.}
\label{fig2}
\end{figure}

In fig.\ \ref{fig1}, $T = 100$ is sufficiently long so 
that $\langle V_x\rangle$ and $\langle V_y\rangle$ relax to $V_{NS}(c)$ 
during the green phase of the corresponding signals and to zero during 
the red phase of the signals. In contrast, for the same values 
of the parameters $D$ and $c$, 
the average speeds $\langle V_x\rangle$ and $\langle V_y\rangle$ do not 
get sufficient time 
to relax to zero during the red phase of the corresponding signal 
provided $T$ is sufficiently small, e.g., $T = 20$ (see fig.\ \ref{fig2}). 

\begin{figure}[hbt]
\epsfxsize=\columnwidth\epsfbox{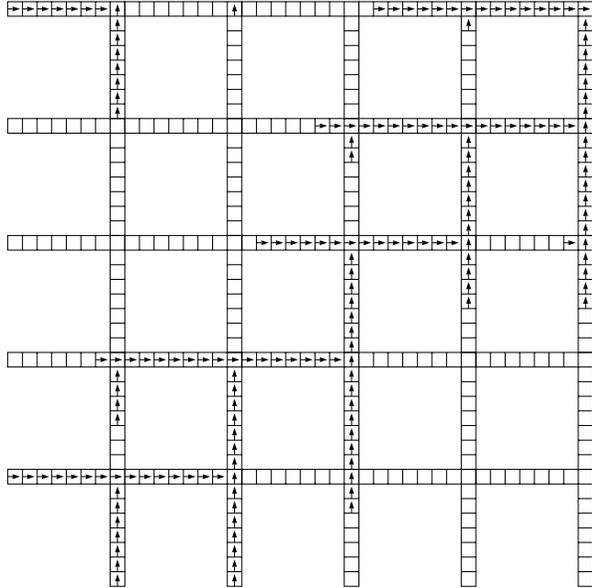}
\caption{A typical jammed configuration of the vehicles. 
The east-bound and north-bound vehicles are represented by the 
symbols $\rightarrow$ and $\uparrow$, respectively.}
\label{fig3}
\end{figure}

The most dramatic result of our investigation is that at a 
sufficiently large density $c_*(D,T)$, which depends on $D$ and $T$, a 
phase transition from the "free-flowing" dynamical phase to 
a completely jammed phase can take place in our "unified" model. 
In the jammed phase, the flow of east-bound 
vehicles is blocked by the north-bound vehicles and vice-versa; 
this gridlock phenomenon as well as the typical configurations 
in the jammed phase (see fig.\ \ref{fig3}) of our "unified" 
model are similar to those in the BML model. In spite of these 
apparent similarities, as we shall explain now, the mechanism 
that triggers jamming in our "unified" model is different from 
that in the BML model. It is obvious from the updating rules 
that if $p = 0$, i.e., if no random braking takes place, complete 
jamming is impossible in this model at any density $c < 1$.
Therefore, a vehicle which is located 
at the crossing of two mutually perpendicular streets and whose 
instanteneous speed is $V = 1$ at the end of the deceleration 
stage (i.e., having at least one empty site in front of it) 
would vacate the crossing unless its speed is reduced to $V = 0$ 
because of random braking. However, if such a halt at a crossing 
due to random braking takes place at the last time step before 
the signal for it turns red, it would not only continue to block 
the perpendicular flow of traffic through the same crossing during 
the next $T$ time steps but would also give rise to a queue of 
jammed vehicles in the perpendicular street passing through the 
same crossing. Therefore, when $p \neq 0$ and the density is 
sufficiently high, the dynamical phase of "freely-flowing" traffic 
becomes unstable against the spontaneous formation of jams and the 
entire traffic system self-organizes so as to reach the completely 
jammed state. 

\begin{figure}[hbt]
\epsfxsize=\columnwidth\epsfbox{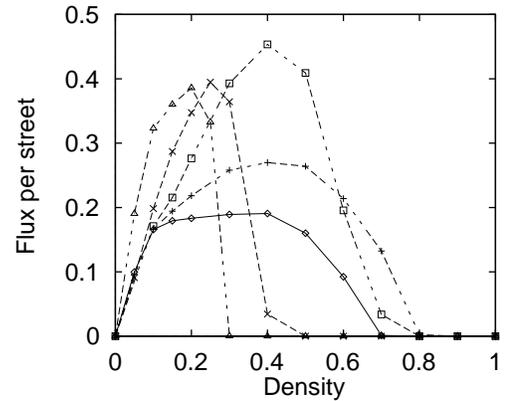}
\caption{The fundamental diagram. The symbols
$\diamond, +, \square, \times$ and $\triangle$ correspond, respectively, 
to $T = 100, 50, 20, 10, 4$. The common parameters are 
$V_{max} = 5, p = 0.5$, $L=100$, and $D = 20$.}
\label{fig4}
\end{figure}

Moreover, for given $D$, the shorter is the time interval $T$ 
the smaller is the $c_*$ (see fig.\ \ref{fig4}). Besides, the density 
corresponding to the maximum flux also shifts to smaller densities 
with the decrease of $T$. Furthermore, the maximum throughput 
is a non-monotonic function of $T$ in the "free-flowing" phase; 
this result may be of practical use in traffic engineering 
for maximizing the throughput. 

In this communication we have developed a ``unified'' model where
the jams are created by the same stochastic process as in the NS model, 
but the transition to complete jamming and the jammed configurations 
are very similar to those in the BML model. 
We have also established that our 
model describes the time-dependence of the average speeds of the 
vehicles in more realistic manner than any of the earlier CA 
models of BML-type. Results of our ongoing investigations of the 
effects of (a) turning of vehicles from east-bound (north-bound) 
streets to north-bound (east-bound) streets and (b) green-wave 
signalling on the flow and jamming 
will be reported elsewhere ~\cite{chowsh}. 

We thank Ludger Santen for useful discussions as well as for help in 
producing figure\ \ref{fig3} and Dietrich Stauffer for valuable comments 
on the manuscript. This work is supported in part by the SFB341 
Aachen-J\"ulich-K\"oln.

\end{document}